# SnAs: a 4K weak type-II superconductor with non-trivial band topology

M.M. Sharma[1,2], N.K. Karn[1,2], Prince Sharma[1,2], Ganesh Gurjar[3], S. Patnaik[3] and V.P.S. Awana[1,2, *]

[1] *Academy of Scientific & Innovative Research (AcSIR), Ghaziabad-201002*
[2] *CSIR- National Physical Laboratory, New Delhi-110012*
[3] *School of Physical Sciences, Jawaharlal Nehru University, New Delhi-110067*

**Abstract**

Superconductors with non-trivial band topology are emerging as one of the best avenues to study quantum anomalies and experimental realization of Majorana Fermions. This article reports the successful crystal growth of superconducting SnAs, which can have topologically non-trivial states, as evidenced in DFT (Density Functional Theory) calculations, Z2 invariants and topologically surface state. Here, we followed a two-step method to grow SnAs crystal. The powder XRD (X-ray Diffractometry) pattern of synthesized crystal ensures that the crystal is grown in a single phase with a NaCl type cubic structure and the EDAX (Energy Dispersive X-ray Analysis) endorses the stoichiometry of the as-grown sample. The DFT calculations performed with and without the inclusion of spin-orbit coupling (SOC) show band inversion at various K symmetry points near the Fermi level. The recorded Raman spectra show two different modes, which are assigned as $A_1$ and $E_{TO}$ vibrations. The ZFC (Zero-Field Cooled) & FC (Field Cooled), as well as the isothermal M-H (Magnetization vs. field) measurements, are also performed for the topological non-trivial SnAs superconductor, which eventually confirm the weak type-II superconductivity at 4K. Various other superconductivity parameters viz. kappa parameter, coherence length, and penetration depth are also calculated to probe the as-grown sample's characteristics.

*Corresponding Author

Dr. V. P. S. Awana:  E-mail: awana@nplindia.org
Ph. +91-11-45609357, Fax-+91-11-45609310
Homepage: awanavps.webs.com

**Introduction:**

Non-trivial band topology creates opportunities for condensed matter scientists to search for the materials with such kind of band structures. In this regard, Bi rich materials viz. $Bi_2Se_3$, $Bi_2Te_3$, $Bi_{1-x}Sb_x$ are studied to a greater extent as these materials are classified as topological insulators (TIs) [1-3]. The TIs are found to show fascinating properties such as persistence of giant MR [4], dissipation less surface current [5], superconductivity on doping with adequate dopant [6-8], generation of THz [9,10] and thermoelectric properties [11], etc. Superconducting TIs create a new class of materials named as Topological Superconductors [TSCs] [12]. The TSCs are important, as these are auspicious materials for the experimental realization of Majorana Fermions [13]. A TSC consists of a bulk superconducting gap along with topologically non-trivial surface states [12]. Apart from possible experimental realization of Majorana fermions, TSCs also host some exotic properties such as the Anomalous Josephson effect [12, 14], quantized thermal conductivity [15] and zero bias energy modes [16] etc. Topological superconductivity can be appreciated in some doped TIs [6-8], while there are some known superconductors [17,18] that also show topological superconductivity in their pristine form.

Apart from Bi rich materials, some Sn-based system such as SnTe [19], SnSb [20], and SnAs [21] also supposed to have topological non-trivial surface states. SnSb and SnAs are superconducting in their intrinsic form and can be gazed at as possible candidates of topological superconductors [22,23]. Both of these materials were reported to be superconducting in a combined report on Sn-based superconductors [23]. In SnAs, the valence state of Sn remained ambiguous for a long time; it was unclear whether Sn is in a single valence state of $Sn^{+3}$ or it is in a mixed-valence state of $Sn^{+4}$ and $Sn^{+2}$. It was suggested that superconductivity in SnAs is related to mixed-valence states of Sn [24], but this was utterly rejected in a later report [25]. HAXPES (Hard X-Ray Photoelectron Spectroscopy) measurements were performed on SnAs that showed Sn to be in single valence state, i.e., in Sn + 3 state [25]. It is also expected that SnAs undergo a structural phase transition from NaCl type structure to CsCl structure by application of pressure [26], and $T_c$ can be enhanced threefold to up to 12.2K [27]. Band dispersion of SnAs pretty much resembles that of SnTe, a topological crystalline insulator, creating a possibility that SnAs also have non-trivial band topology. A recent report on the DFT (Density Functional Theory) based band structure calculations of SnAs [28] showed little effect of SOC (Spin Orbit Coupling). On the other hand, considerable splitting in band structure is established in ARPES (Angle Resolved

Photo Electron Spectroscopy) spectra [28]. In fact, there are only few reports existing on the superconducting and topological behavior of SnAs, and that also not agreeing completely with each other. The discrepancy in valence state and possible non-trivial band topology along with superconductivity encourages one to study further SnAs.

In the present article, we proposed a two-step synthesis route for crystal growth of SnAs The XRD and EDAX analysis endorsed the single-phase crystallization and stoichiometry of the as-grown crystal. There is no sign of secondary phase distortion in the Rietveld refinement of the XRD pattern. Raman spectra of SnAs is not reported until now as per our knowledge; we show three Raman peaks in recorded Raman Spectra, which are similar to those observed for its isostructural compound like SnTe [29]. The FC (Field Cooled) – ZFC (Zero Field Cooled) measurements confirm the bulk superconductivity near 4K, while the isothermal magnetization (M-H ) plots endorse a weak type-II superconductor. DFT calculations are also carried out to know the impact of SOC on the band structure of SnAs, which showed a band inversion as well as the introduction of gapped states near the Fermi level. These calculations indicate the possibility of the existence of topological surface states in SnAs superconductor.

**Experimental:**

Single crystalline SnAs sample was grown by following a two-step method. In the first step, polycrystalline SnAs were prepared using high purity (4N) powders of Sn and As in the stoichiometric ratio. These powders were ground thoroughly with Agate Mortar pestle in an MBRAUN glove box filled with Argon gas, and thus, obtained homogenous powder was palletized by using a hydraulic press. This palletized sample was then vacuum evacuated in quartz ampoule with a pressure of $5*10^{-5}$. Then the sample is heated to $500^0$C at a prolonged rate of $30^0$C/h and kept at this elevated temperature for 48 hours after that it is allowed to cool naturally. Thus obtained sample was reheated to $550^0$C for 48 hours following the same vacuum encapsulation process. The image of this vacuum evacuated polycrystalline sample is shown in the left inset of Fig. 1. In the second step, the vacuum evacuated polycrystalline sample was heated to $900^0$C at a rate of $120^0$C/h. It is kept at $900^0$C for 48 hours to homogenize the sample. Further, the sample is slowly cooled at a rate of $3^0$C/h up to $550^0$C and annealed at this temperature for 48 hours and then allowed to cool generally to room temperature. Thus, obtained crystal of SnAs looks silvery shiny and easily cleavable, as shown in the right inset of Fig. 1. The schematic of this complete heat treatment is shown in Fig. 1.

Powder XRD (PXRD) pattern is taken on the gently crushed powder of synthesized SnAs crystal with the help of the Rigaku Miniflex II X-Ray diffractometer equipped with CuK$_α$ radiation of 1.5418Å wavelength. Raman Spectra of synthesized SnAs crystal is recorded with the help Renishaw inVia Reflex Raman Microscope equipped with Laser of 514nm. & 720nm. The compositional analysis and surface morphologies were studied using the EDAX and scanning electron microscopy (SEM) with Zeiss EVO-50, respectively. Quantum Design PPMS is used for magnetization measurements. Full proof software is used for Rietveld refinement of the PXRD pattern, and VESTA software [30] is used to draw unit cells of the synthesized crystal. Quantum Espresso software is used for DFT calculations which is further extended to investigation of the structure at energy surfaces near fermi level and calculate Z2-invariant to classify the topology present in the system. For this, the first principal methods based on DFT is implemented in QUANTUM ESPRESSO to obtain the electronic band structure and density of states (DOS). Moreover, in order to calculate the Z2 invariants of the SnAs, *WANNIER90* is used in which wannierization of Bloch wave function is implemented [31]. In this regard, the MLWFs (Maximally localized Wannnier Function) are considered, which is used to verify the band structure that is generated through the first principal method. Based on the MLWFs, we obtain an effective tight-binding (TB) model for the system SnAs. This TB model is further processed and implemented in WANNIER-TOOLS [32] on a 8×8×8 K-mesh. The energy surface near fermi level is obtained using WANNIER-TOOLS. The Wannier charge centers are further calculated where the evolution of Brillouin zone planes is studied which indicates the states of Z2-invariant. Fermi surface and First Brillouin zone are drawn by using Xcrysden software, for Fermi surface data has been taken from WANNIER TOOLS.

**Results & Discussion:**

The XRD pattern of the as-grown crystal is performed, which confirmed the single-phase formation of the crystal. Rietveld refined PXRD pattern of synthesized SnAs crystal is shown in Fig. 2(a). Rietveld refinement ensures that the sample is crystallized in cubic NaCl type structure with the Fm-3m space group. All XRD peaks are labeled with their respective planes, as shown in Fig.2(a). No impurity peak is observed, discarding the possibility of any secondary phase in the synthesized sample. Fitting parameters that offer the quality of fit being $\chi^2$ (goodness of fit), $R_p$, and $R_{wp}$, are found to be 3.50, 10, and 13.3, respectively. All these values are accep30table for showing the quality of fit and the excellence of synthesized crystal. Lattice parameters obtained from Rietveld analysis are a = b = c = 5.721(1)Å and $\alpha = \beta = \gamma = 90^0$. The atomic positions are found to be Sn (0.5, 0.5, 0.5) and As (0, 0, 0). Based on the XRD pattern, it is clear that no detectable structural distortion is present in the synthesized SnAs sample as all observed peaks are well fitted in the Fm-3m space group. The synthesized crystal's unit cell is drawn using VESTA software, see Fig. 2(b). It indicates that each Sn atom is connected to six As atoms, and the synthesized sample has a NaCl type structure. Fig. 3 shows the EDAX spectra of the synthesized SnAs sample, which shows peaks for the constituent elements viz. Sn and As without any impurity. The stoichiometry of the crystal is also near stoichiometric. The inset of Fig. 3 shows a scanning electron microscopy image of a synthesized SnAs sample taken with a resolution of 10μm. It shows the terrace type morphology, typical for crystalline samples, and confirms the crystalline nature of the synthesized SnAs sample.

Fig. 4(a) shows Raman spectra recorded on mechanically cleaved crystal flake of synthesized SnAs crystal. The sample was irradiated with a laser of 514nm, which is maintained at a low power of 5mW for 30 seconds to avoid sample heating. Recorded Raman spectra show one peak at 70.58cm$^{-1}$ and a broad hump that is deconvoluted into two peaks at 140.81 and 166.04cm$^{-1}$ using the Gauss fitting. There is no report available showing Raman spectra of SnAs to best of our knowledge. Here we compared Raman spectra of SnAs with that of its isostructural compound i.e., SnTe. The SnTe is reported to have two Raman active modes viz. $A_1$ and $E_{TO}$ around 120cm$^{-1}$ and 140cm$^{-1}$ due to the Te environment around Sn, in a NaCl type structure [29]. Here in SnAs, Te is a compliment to As that resulted in shifting of Raman peaks towards higher frequency, as the same depend on $(k/\mu)^{1/2}$ where k is force constant and μ is reduced mass. The As atom is smaller than Te and has lesser atomic weight, hence resulting in lowering of reduced mass in SnAs than that in SnTe. This is the reason that Raman shift peaks are observed to be shifted towards higher frequencies at 140.81 and 166.04cm$^{-1}$ in SnAs. These Raman modes are consistent with theoretical studies made on phonon modes of SnAs [33]. The schematic of $A_1$ and $E_{TO}$ vibrations in SnAs are shown in Fig. 4(b). Raman modes observed at 70.58cm$^{-1}$ and 166.04cm$^{-1}$ are labeled as $E_{TO}$ vibrational modes, which are doubly degenerate. In these vibrations, all the As atoms vibrate in and out of the plane directions. Simultaneously, the $A_1$ vibrational mode is observed at 140cm$^{-1}$, in which the As atoms that are in the plane of Sn atoms, show only in-plane vibrations. In contrast, the As atoms in the vertical direction to Sn atoms show out of the plane movement.

In general, the materials having Fm-3m symmetry or NaCl type structure are not supposed to have Raman active modes [34]. In fact, this type of structure consists of two interpenetrating atomic lattices in which lattice composed of one kind of atom vibrates against the other and hence finally making the vibrations antisymmetric with respect to the center of symmetry. This makes the material Infra-Red (IR) active but Raman inactive. In SnTe also, Raman modes are observed after slight doping [29] or at low temperatures after a structural transformation [35]. The presence of three Raman shift peaks in recorded Raman spectra indicates that there may be slight distortion in studied SnAs crystal structure, which can be so small that it cannot be detected in XRD. It is thus warranted that the present SnAs crystal is subjected to synchrotron radiation experiments to resolve its perfect structure.

Fig. 5 shows DC magnetization measurements of synthesized SnAs crystal under FC & ZFC protocols in the presence of a magnetic field of 12Oe. A strong diamagnetic signal is observed with $T_c^{onset}$ near 4.2K in both FC and ZFC measurements, which is higher than previously observed $T_c$ of SnAs samples [25,28]. The diamagnetic signal starts to saturate near 3.8K, showing a transition width of about 0.4K. The shielding (ZFC) and Meissner (FC) superconducting volume fractions below diamagnetic saturation are calculated to be around 35% and 20% respectively. The difference between FC signal and ZFC signal points out the possible flux pinning in the synthesized SnAs crystal. It signifies that the observed superconductivity in synthesized crystal is of weak type-II in nature. Similar kind of FC and ZFC signals has recently been observed in SnTaS$_2$ [36], which is also a type-II superconductor. Further M-T measurements were carried out at different fields viz. 12Oe, 30Oe, 50Oe, 100Oe, 200Oe and 400Oe to examine the type-II behavior of synthesized crystal and the same is shown in Fig. 6(b). It can be clearly seen in Fig. 6(b) that $T_c$ is decreasing as the applied field is increased. $T_c$ is observed to be 4.2K, 3.9K, 3.8K, 3.4K and 2.9K at applied fields of 12Oe, 30Oe, 50Oe, 100Oe and 200Oe respectively. No superconducting transition is observed at 400Oe. It is clearly visible in Fig. 6(b) that the diamagnetic signal is sharp upto 50Oe while it becomes broader at 100Oe further the ZFC signal does not saturate up to 2K. Broadening of diamagnetic transition at 100Oe suggest that the synthesized SnAs crystal is in its mixed state and the non-saturating diamagnetic transition at 200Oe confirms that superconductivity persists at 200Oe. In previous report on same material, it was shown that the critical field ($H_c$) of SnAs is around 130Oe [25,28]. These results show that studied SnAs crystal is a type-II bulk superconductor 4K.

Fig. 6(a) shows isothermal magnetization (M-H) plots of synthesized SnAs crystal at various temperatures viz. 2.0, 2.2, 2.4, 2.6, 2.8, and 3.0K. It can be seen that critical fields viz. lower critical field ($H_{c1}$) and upper critical field ($H_{c2}$) are seen decreasing as the temperature is increased towards the normal state. These M-H plots further indicate that the synthesized SnAs crystal is a weak type-II superconductor, which contrasts with the reported type-I superconductivity for the same [25]. Interestingly, type-II superconductivity was also proposed in only another report on SnAs superconductivity [28], based on small hysteresis in the M-H loop and presence of differential paramagnetic in FC measurements, signifying the presence of a mixed or vortex state. This makes type of superconductivity in SnAs quite ambiguous. Nature of superconductivity in presently studied SnAs becomes more explicit in Fig. 6(b), which depicts the M-H loop of synthesized SnAs crystal at 2K. In this Fig., the $H_{c1}$ and $H_{c2}$ are marked with an arrow as $H_{c1}$ is being defined as the field at which the M-H loop deviates from linearity, and $H_{c2}$ is defined as the field where the M-H loop closes and starts to touch the baseline for a type-II superconductor. In Fig. 6(b), it can be seen that the M-H loop starts to deviate from linearity roughly from 60Oe (a more accurate value of $H_{c1}$ will be calculated in later part), showing $H_{c1}$ to be near to 60Oe and this M-H loop closes at 291Oe, indicating $H_{c2}$ to be 291Oe. This value of $H_{c2}$ is nearly 5 times larger than $H_{c1}$, clearly showing that the synthesized SnAs sample is a type-II superconductor. Further, it is a weak type-II superconductivity as the value of $H_{c2}$ is not large enough. Inset of Fig. 5(b) shows the M-H loop at 3.6K; it is interesting to see that the M-H loop is still open even at 3.6K, which is considered the $T_c$ of SnAs according to previous reports [23,25,28]. This M-H loop shows that type-II superconductivity in synthesized SnAs crystal persists up to 3.6K.

It is a tedious task to determine the correct value of lower critical field $H_{c1}$, as it is the field at which a superconductor enters into a mixed state or the field at which just one magnetic line of force penetrates the superconducting sample. There are several methods to determine $H_{c1}$ by different groups [37,38]. Here we followed the method proposed in ref. 37, in this method, the motive is to determine the point from where the M-H loop transits from linear behavior to nonlinear one. For this, we determined the slope of the linear fit of the M-H loop at low fields, and then this slope is used to calculate $M_0$. Thus, obtained values of $M_0$ are then subtracted from each isotherm to get $\Delta M$ and then plotted against the applied field, see Fig. 6(c). A baseline is created at $\Delta M=0$, $H_{c1}$ is the point at which the $\Delta M$ vs. H plot deviates from this zero baseline. $\Delta M$ vs. H plot at 2K is shown in the upper inset of Fig. 6(c) in which $H_{c1}$ is marked with an arrow pointing the deviation of the plot from zero baselines. The obtained values of $H_{c1}$ from this method are 55Oe, 48Oe, 41Oe, 34Oe, 28Oe, and 22Oe at 2, 2.2, 2.4, 2.6, 2.8, and 3.0K, respectively. Further the value of upper critical field $H_{c2}$ are determined as the field at which M-H loop closes. The values of $H_{c2}$ are found to be 290Oe, 268Oe, 250Oe, 213Oe, 183Oe and 147Oe at 2, 2.2, 2.4, 2.6, 2.8, and 3.0K respectively. These $H_{c1}$ and $H_{c2}$ values are plotted against their respective temperature, and the plot is fitted with quadratic G-L equations viz. $H_{c1}(T) = H_{c1}(0) [1-T^2/T_c^2]$ and $H_{c2}(T) = H_{c2}(0) [1-T^2/T_c^2]$. This exercise is performed to draw superconducting phase diagram of synthesized SnAs crystal and shown in Fig. 6(d). Superconducting state, mixed state and normal state are clearly visible in phase diagram which also shows type-II superconductivity of studied crystal. From this phase diagram, the values of $H_{c1}(0)$ (lower critical field at absolute zero) and $H_{c2}(0)$ (Upper critical field at absolute zero) are determined and the same are found to be 81Oe and 474Oe.

From the M-H loop at 2K, we found the upper critical field's value to be 291Oe, and the value of lower critical field $H_{c1}$ as 55Oe. The mean critical field of a type-II superconductor can be calculated using the formula $H_c = (H_{c1}*H_{c2})^{1/2}$, and is calculated to be 126.51Oe. The upper critical field at absolute zero $H_{c2}(0)$ can be calculated by using the Ginzberg-Landau (G-L) equation i.e.

$$H_{c2}(T) = H_{c2}(0) * \left[\frac{1-t^2}{1+t^2}\right]$$

Where $t = T/T_c$ and called a reduced temperature, T is the temperature for which upper critical field value is used, i.e., 2K and $T_c$ is critical temperature, which is taken to be 4K as observed in FC and ZFC measurements, giving reduced temperature to be 0.5. Following this equation, the value of $H_{c2}(0)$ is found to be 485Oe which is very close to that observed from phase diagram. The Ginzberg-Landau (G-L) kappa ($\kappa$) parameter is calculated with the help of equation $H_{c2}(0) = \kappa*(2)^{1/2}*H_c$, and is found to be 2.71, which is above the threshold value of type-I superconductivity of 1/(2)$^{1/2}$. This value of κ parameters confirms type-II superconductivity in presently synthesized SnAs crystal. Another superconductivity critical parameter coherence length ξ(0) can be calculated using the formula $H_{c2}(0) = \frac{\varphi_0}{2\Pi\xi(0)^2}$, where φ$_0$ is flux quanta, and its value is 2.0678 x 10$^{-15}$Wb. The value of ξ(0) is found to be 8.25Å. From the value of ξ(0) and κ, the value of penetration depth λ(0) can be calculated by using the relation κ = λ(0)/ξ(0), this is found to be 22.35Å. The physical parameters of synthesized SnAs crystal are given in Table-I.

Bulk electronic band structure and Density of States (DOS) of synthesized SnAs crystal are theoretically studied within the etiquettes of Density Functional Theory (DFT). The crystal parameters are considered from the Rietveld refinement analysis to calculate the band structure theoretically. The calculation assesses the spin-orbit coupling (SOC) and without SOC as implemented in Quantum Espresso with Perdew-Burke-Ernzerhof (PBE) exchange-correlation functional [39,40]. Fig. 7(a) is showing the DFT calculated Fermi surface of synthesized SnAs crystal which is calculated using WANNIER90. Bulk electronic band structure is calculated along the K-path X → W → L → Γ → K calculated from the SeeK-path: the k-pathfinder and visualizer [41]. This particular path is shown in Fig. 7(b) where the First Brillouin zone indicates the K-path, that is chosen for the calculations of bulk electronic band structure. The bulk electronic band structure (left hand side image) along with DOS (right hand side image) without (w/o) SOC and with SOC of synthesized SnAs crystal is shown in fig.7(c).

In fig.7(c), the bulk electronic band structure w/o SOC are shown through fine lines while the band structure that were calculated by considering SOC is shown through the thicker lines. A Dirac point like bands near L point has been clearly shown in the w/o SOC band structure, however, it completely disappears with the inclusion of SOC. This eventually suggests the occurrence of band inversion phenomenon with the inclusion of SOC. It signifies that SOC is effective in the studied system. Similar type of gap in band structure is also shown to be existed with the SOC in some previous reports [21,28]. Moreover, the bands have six-fold degeneracy at the Γ point in w/o SOC plots. This six-fold degeneracy has been raised and the bands have become four-fold degenerate when SOC is included as shown in inset of Fig. 7(c). This claims the presence of band degeneracy and it is being lifted, when SOC is included, this also confirms the effectiveness of SOC parameters on bulk electronic band structure of studied system. On the other hand, SOC is found to have negligible impact on DOS, which suggest that SOC does not affect superconducting properties of SnAs. Moreover, DOS does not vanish at the Fermi level which signifies the metallic or semi-metallic character of studied system. It is correspondingly confirmed through bulk electronic band structure as bands are found to cross the Fermi level.

Further, the fermi energy surface projected in (110) plane ($k_x k_y$-plane) is also calculated. The $\Gamma$ point is chosen as the center of the $k_x k_y$-plane in order to calculate the states as shown in fig.7(d). The lower energy surface corresponds to the band crossing Fermi-level in electronic band structure. While the upper energy surface is for the lowest unoccupied band. It is found through the calculation that the lower energy surface center is a saddle point while, it is a stable point with Dirac type cone shape in the upper surface region. It suggests that the studied systems contain topological surface states too, which is in consistent with the previous report on the same material in ref. 21. The presence of topological surface states along with non-vanishing DOS at the Fermi level suggest that the studied system can be classified as topological semimetal.

The first principal calculation of band structure shows that the system respects the Time-Reversal Symmetry (TRS) since each band splits with the inclusion of SOC. Further, it is well known that Chern number is not the good quantity for TRS systems in order to characterize the topology of a system. Thus, in this study, Z2-invariants are also calculated for SnAs that are more suited for TRS systems [42]. In this regard, the Soulyanov-Vanderbilt [42] method of Wannier Charge Centers (WCC) is considered in which the MLWFs is calculated. These WCC are evolved in the 6-planes – $K_1$, $K_2$, $K_3 = 0$ and $K_1$, $K_2$, $K_3 = 0.5$ in the Brillouin zone. An even number of the crossing of WCC implies a topologically trivial state(Z2=0), whereas an odd number of crossings indicate the presence of a topologically non-trivial state(Z2=1). The Z2 topological number for these six planes are

(a) $k_x = 0.0$, $k_y$–$k_z$ plane: Z2 = 0.

(b) $k_x = 0.5$, $k_y$–$k_z$ plane: Z2 = 0.

(c) $k_y = 0.0$, $k_x$–$k_z$ plane: Z2 = 1.

(d) $k_y = 0.5$, $k_x$–$k_z$ plane: Z2 = 0.

(e) $k_z = 0.0$, $k_x$–$k_y$ plane: Z2 = 1.

(f) $k_z = 0.5$, $k_x$–$k_y$ plane: Z2 = 0

The Z2 topological invariants are shown in Fig. 8. The topological Z2 index is represented as $(v_0; v_1 v_2 v_3)$. The last three Z2 numbers are the weak index which have the Z2 value for $k_i$=0.5 plane (i=x, y and z), whereas the first one is the strong index. The strong index has some redundancy as we can see that for the first pair of the plane – Fig. 8(a) and 8(b), the $v_0 = 0$ indicating topologically trivial, but for the other two pairs of planes, it is $v_0 = 1$, which indicates a topologically non-trivial state. Here the weak index has no redundancy, and it is $(v_0; 000)$. Thus, Z2 calculation shows that SnAs has strong non-trivial topology in $k_y$=0 and $k_z$=0 plane but it is topologically trivial in $k_x$=0 plane. This study not only experimentally shows the superconducting behavior of SnAs, but also proves the topological character through DFT.

**Conclusion:**

We synthesized crystalline SnAs, using a solid-state reaction route and further analyze the structural characteristics and superconductivity of the same. Rietveld refined PXRD pattern showed single-phase growth of SnAs crystal. The Raman spectra of SnAs indicate $A_1$ and $E_{TO}$ modes. FC and ZFC measurements show bulk superconductivity at 4.2K, which is higher than previously reported values. The isothermal M-H plots show weak type-II nature of superconductivity. The DFT calculations confirm the non-trivial band topology in SnAs, as the band inversion is observed with SOC's inclusion. These calculations show the effectiveness of SOC in the system and, thereby, the existence of non-trivial topological states. Calculation of Z2 invariants also confirm topological nature of studied system. These topological surface states and the bulk superconductivity create a possibility that SnAs may be a possible choice to explore more about topological superconductivity.

**Acknowledgment:**

The authors would like to thank Director NPL for his keen interest and encouragement. Authors are also grateful to Mrs. Shaveta for Raman spectroscopy measurements, Dr. J.S. Tawale for SEM and EDAX measurements, and Mr. Krishna Kandpal for vacuum encapsulation of the sample. M.M. Sharma would like to thank CSIR for the research fellowship, and Prince Sharma would like to thank UGC for fellowship support. Both the authors are also thankful to AcSIR for Ph.D. registration.

Table-I

Structural and basic superconductivity parameters of synthesized SnAs crystal:

| Physical Properties (SnAs) | Corresponding Values |
|---|---|
| Crystal Structure & Lattice parameters | Cubic, F m -3 m space group <br> a = b = c = 5.721(1) Å, α = β = γ = 90$^0$ |
| Raman Modes positions | 70.58 cm$^{-1}$, 140.81 cm$^{-1}$, 166.04 cm$^{-1}$ |
| Superconducting critical temperature & Critical Fields | $T_c^{onset}$ = 4.2K, ΔT= 0.4K <br> $H_{c1}$ = 55Oe, $H_{c2}$ = 291Oe & $H_c$ = 126.51Oe at 2K <br> $H_{c1}(0)$ = 81Oe and $H_{c2}(0)$ = 485Oe |
| Coherence length | 8.25 |
| Penetration depth | 22.35 Å |
| G-L Kappa Parameter | 2.71 |

**Figure Captions:**

Fig. 1: Schematic of two-step heat treatment followed for SnAs crystal; left inset shows the image of the polycrystalline sample obtained from the Ist step, and the right inset shows the image of crystal obtained after II$^{nd}$ step.

Fig. 2(a): Rietveld refined PXRD pattern of grown SnAs crystal.

Fig. 2(b): Unit cell of SnAs crystal drawn by using VESTA software.

Fig. 3: EDAX spectra of synthesized SnAs sample inset is showing a SEM image of the same.

Fig. 4(a): Deconvoluted Raman spectra of synthesized SnAs crystal at room temperature.

Fig. 4(b): Representation of $A_1$ and $E_{TO}$ vibrational modes in SnAs crystal.

Fig. 5(a): DC magnetization measurements of synthesized SnAs crystal at 12Oe under FC and ZFC protocols.

Fig. 5(b): FC and ZFC plots of synthesized SnAs crystal at different fields viz. 12Oe, 30Oe, 50Oe, 100Oe, 200Oe and 400Oe.

Fig. 6(a): Isothermal Magnetization vs. the applied field (M-H) plots at temperatures 2.0, 2.2, 2.4, 2.6, 2.8, and 3.0K showing weak type-II superconductivity synthesized SnAs sample.

Fig. 6(b): Magnetization vs. applied field (M-H) plot at 2K with marked $H_{c1}$ & $H_{c2}$, inset shows M-H plot at 3.6K.

Fig. 6(c): ΔM vs. H plot of synthesized SnAs crystal showing deviation of Magnetization from slop of low magnetic field data at a temperature from 2K to 3.0K with a step size of 0.2K, inset shows the same at 2K, marking the deviation from zero to determine $H_{c1}$ at 2K.

Fig. 6(d): Superconductivity phase diagram of synthesized SnAs crystal.

Fig. 7(a): Fermi Surface of synthesized SnAs crystal. (b) First Brillouin zone specifying the k-path selected for bulk band structure calculations. (c) Calculated bulk band structure along with Density of states (DOS) w/o and with SOC under the protocols of Density Functional Theory (DFT), inset is showing zoom view of degenerate bands at Γ point.

Fig. 7(b): The fermi energy surface in the first Brillouin zone projected in kz=0 plane i.e. (110) projection.

Fig. 8: The evolution of Wannier charge centers in the six planes of in Brillouin zone. The Z2 invariants are (a) $k_x = 0.0$, $k_y$–$k_z$ plane: Z2 = 0. (b) $k_x = 0.5$, $k_y$–$k_z$ plane: Z2 = 0. (c) $k_y = 0.0$, $k_x$–$k_z$ plane: Z2 = 1. (d) $k_y = 0.5$, $k_x$–$k_x$ plane: Z2 = 0. (e) $k_z = 0.0$, $k_x$–$k_y$ plane: Z2 = 1. (f) $k_z = 0.5$, $k_x$–$k_y$ plane: Z2 = 0.

Fig. 1

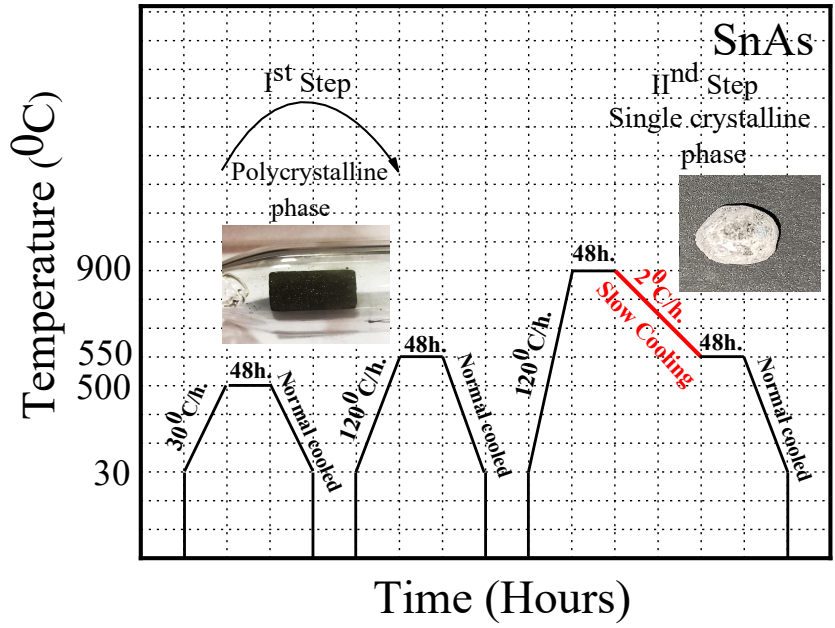

Fig. 2(a)

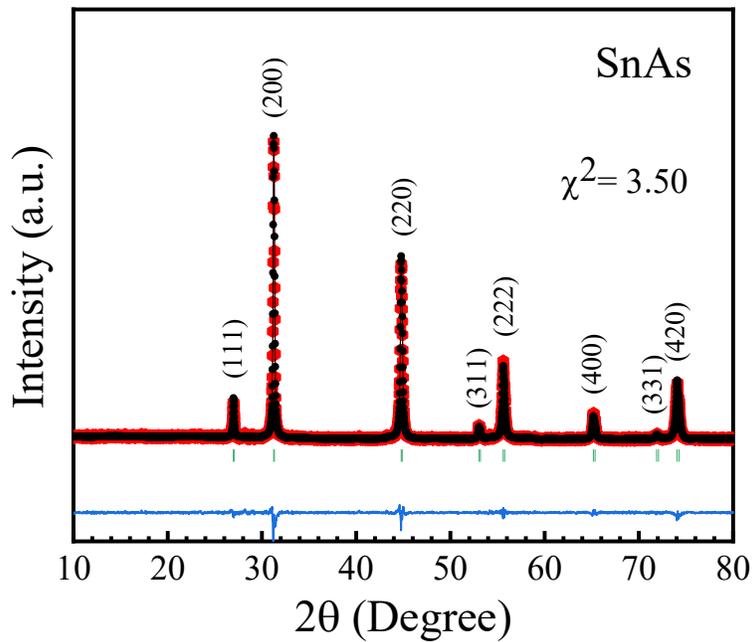

Fig. 2(b)

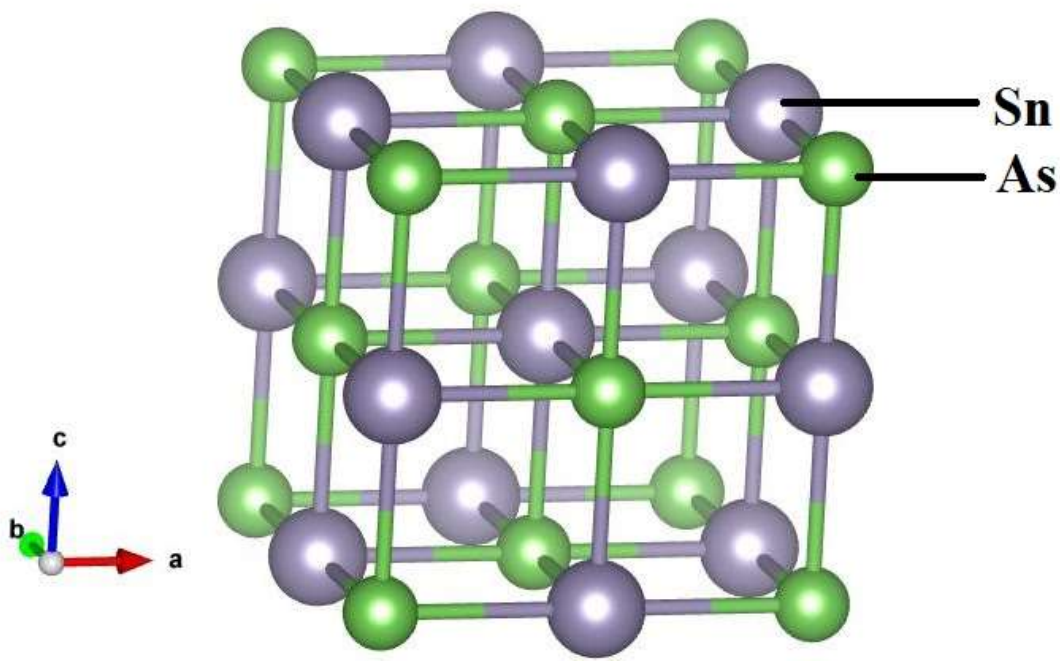

Fig. 3

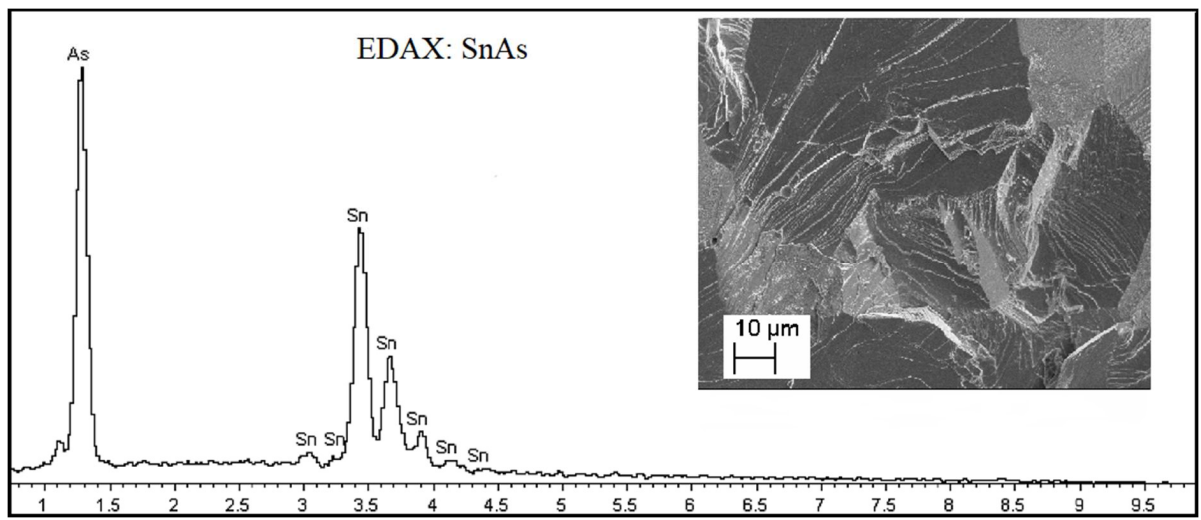

Fig. 4(a)

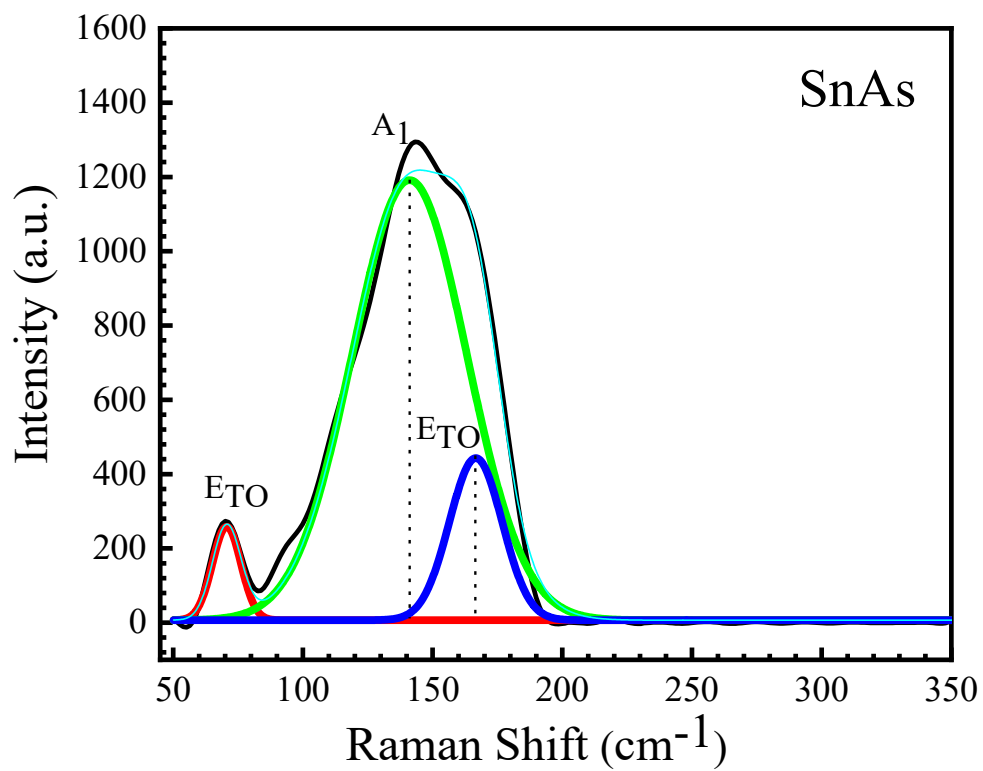

Fig. 4(b)

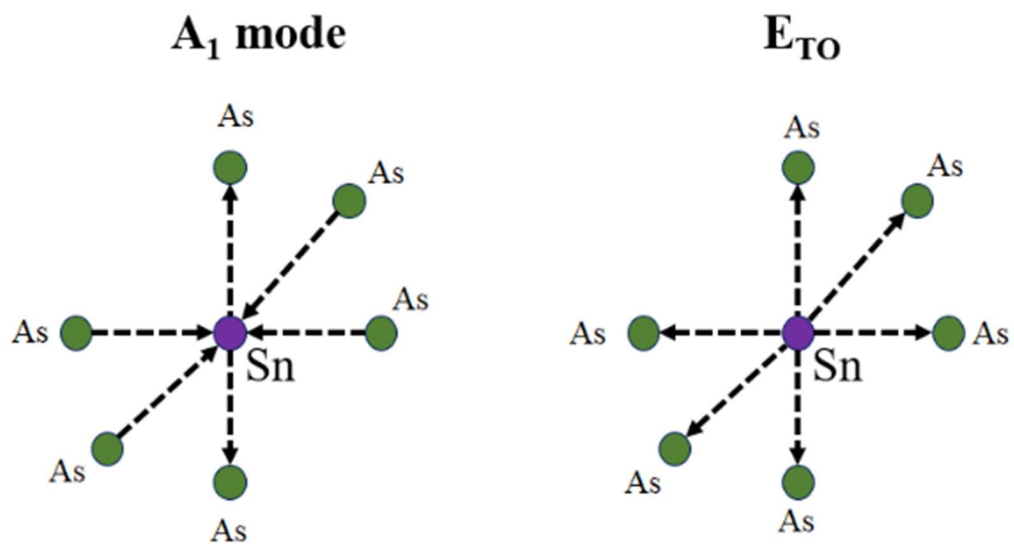

Fig. 5(a)

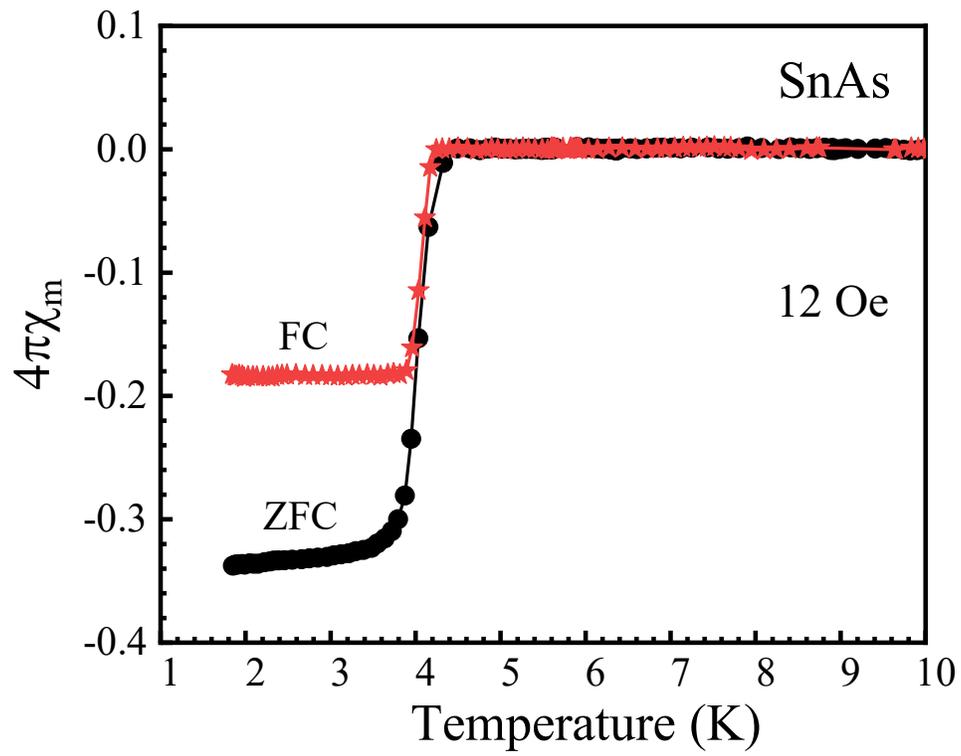

Fig. 5(b)

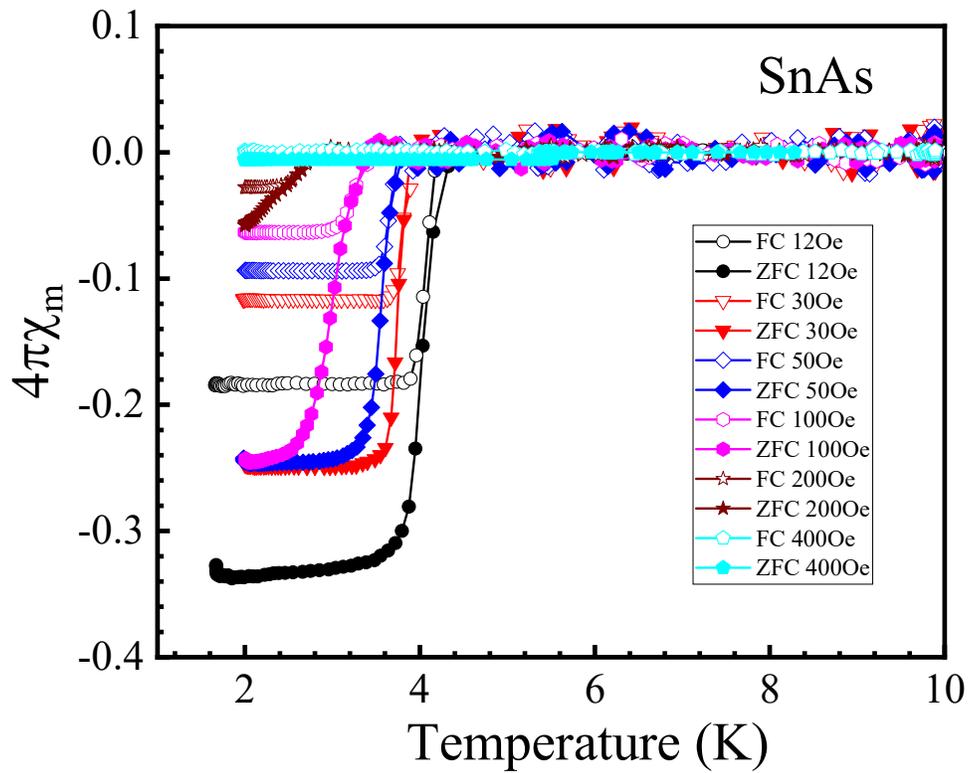

Fig. 6(a)

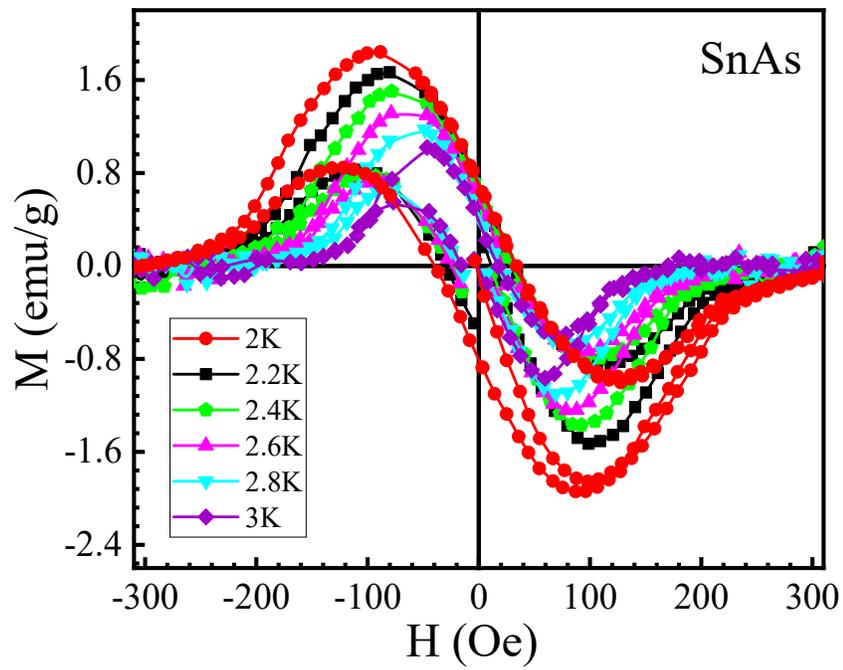

Fig. 6(b)

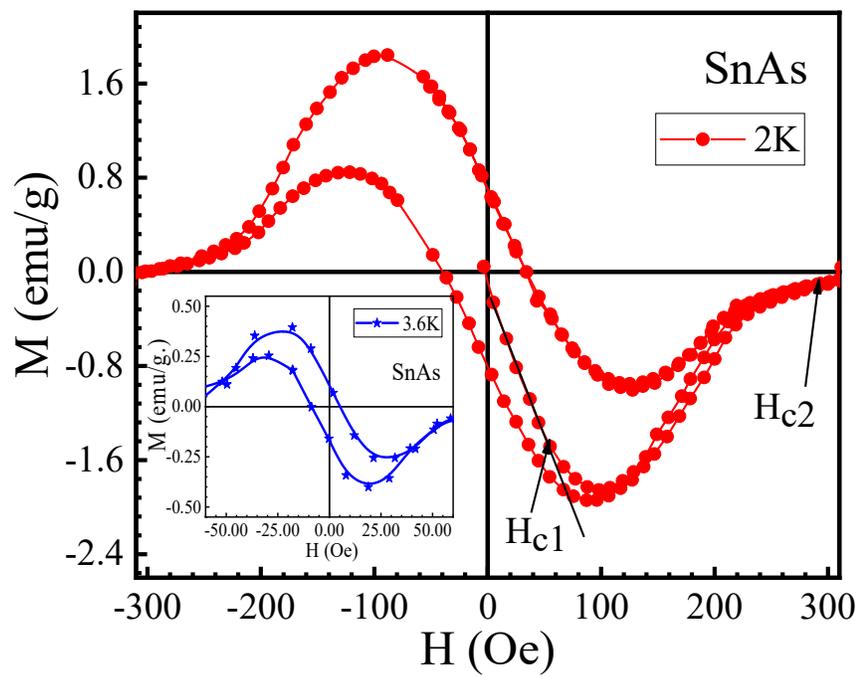

Fig. 6(c)

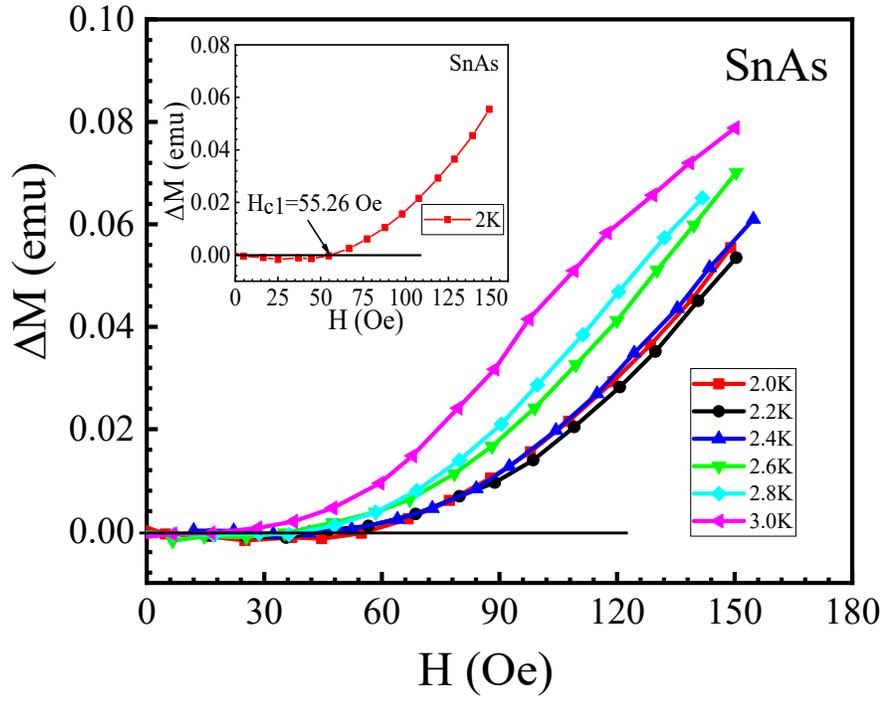

Fig. 6(d)

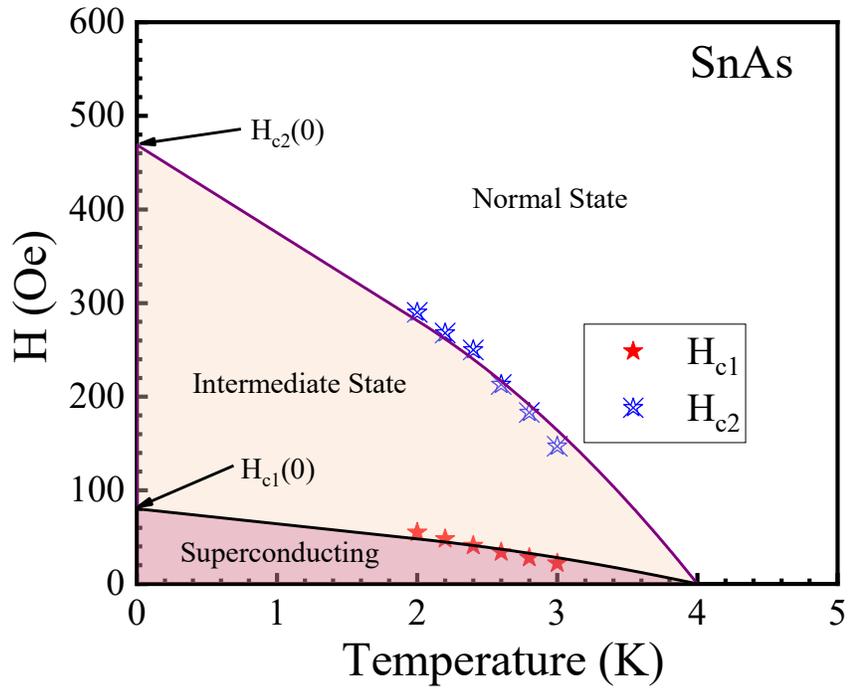

Fig. 7(a)

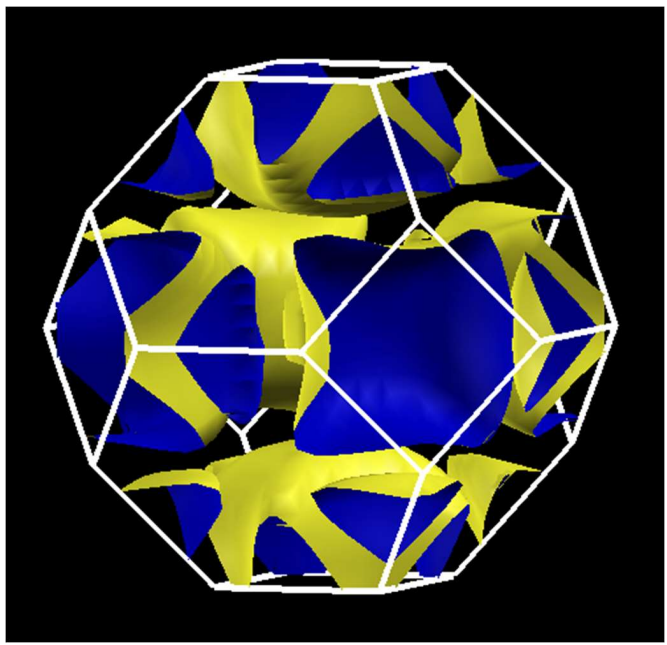

Fig. 7(b)

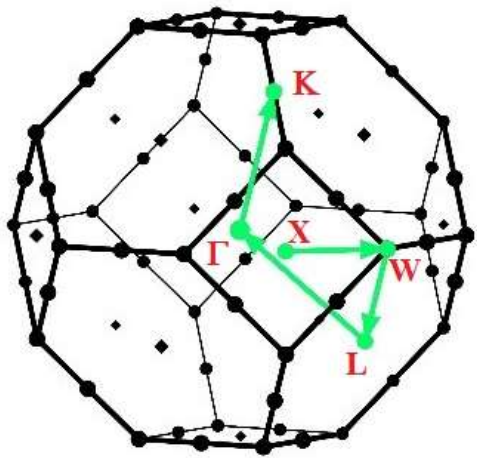

Fig. 7(c)

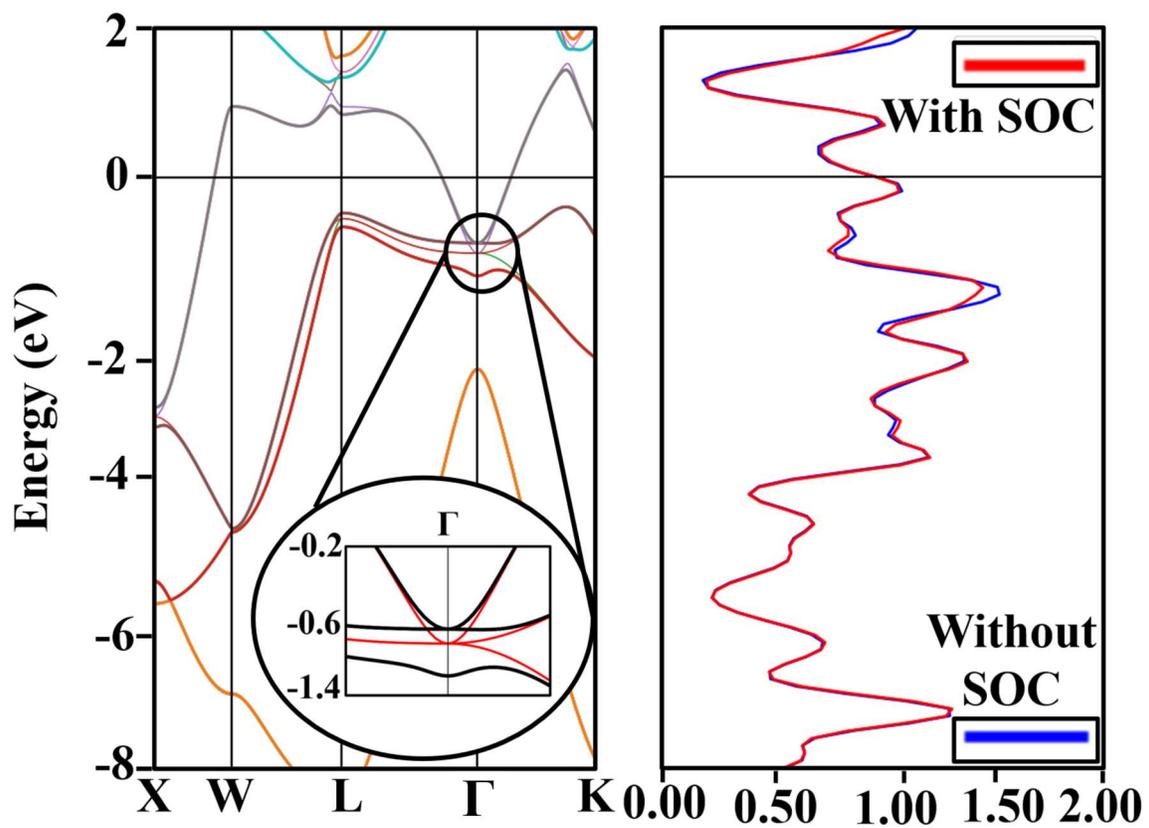

Fig. 7(d)

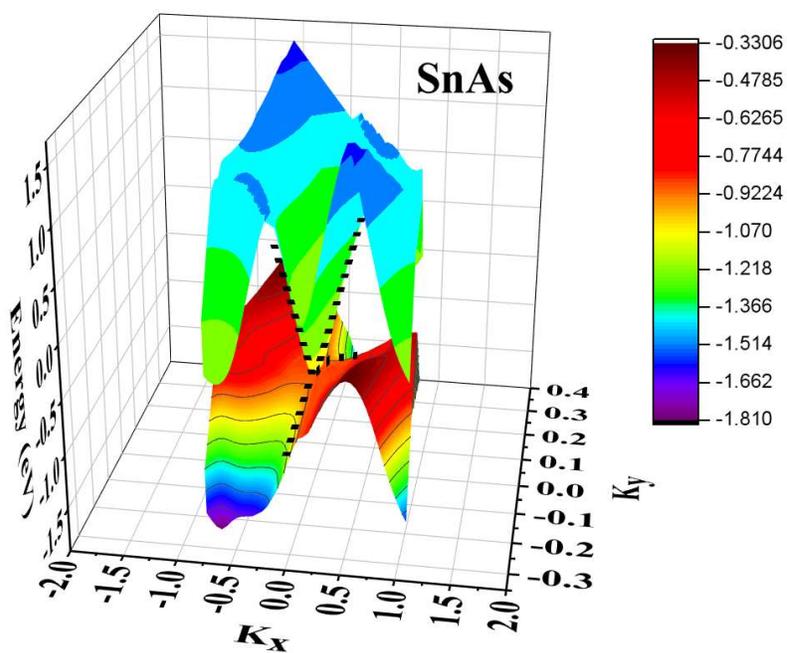

Fig. 8

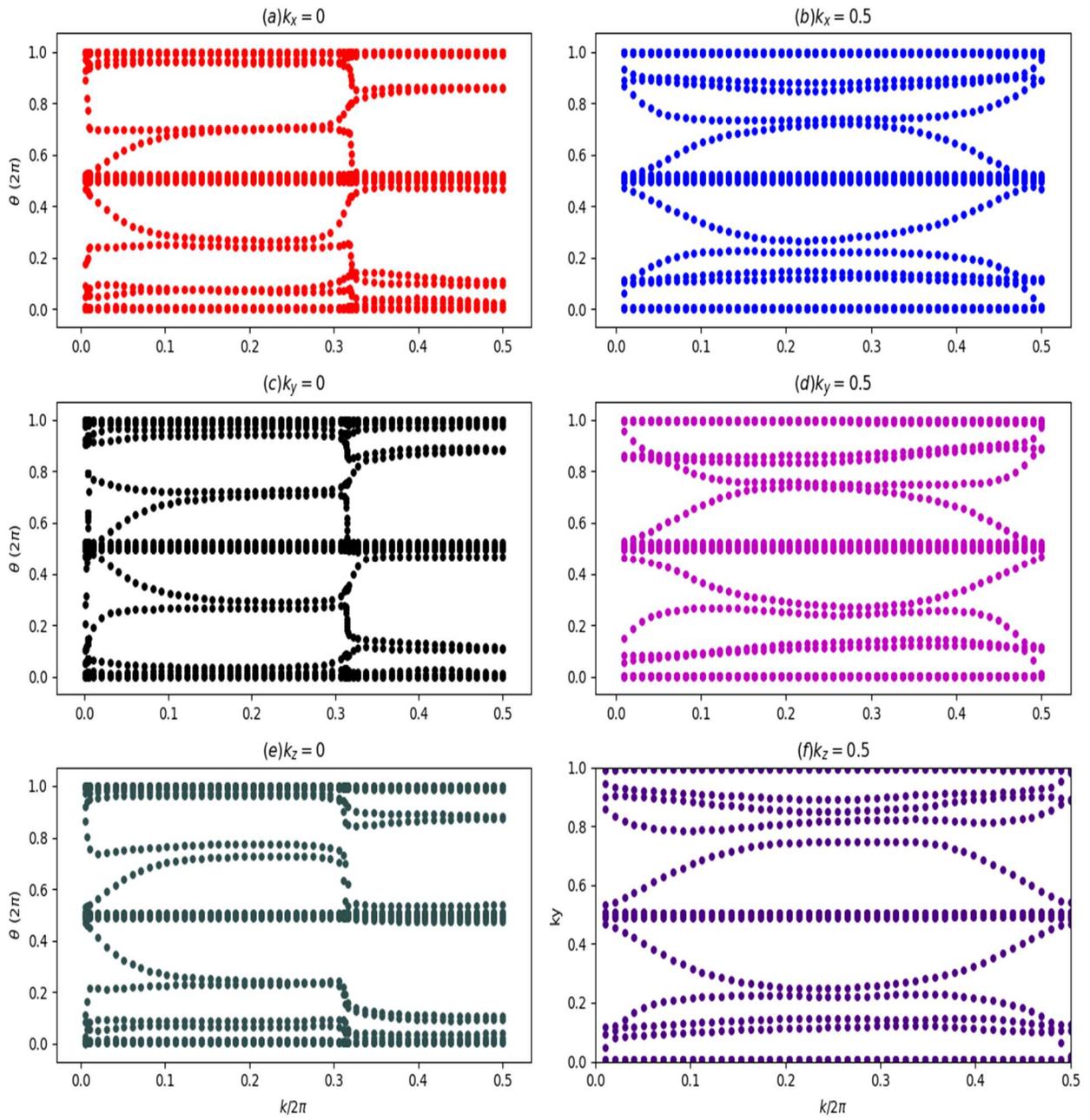